\documentclass[amsmath,amssymb,aps,pra,twocolumn,superscriptaddress,nofootinbib,floatfix]{revtex4-2}

\usepackage{graphicx}
\usepackage{dcolumn}
\usepackage{bm}
\usepackage{siunitx}
\usepackage[colorlinks=true,linkcolor=black,citecolor=blue,urlcolor=black]{hyperref}
\usepackage{comment}
\usepackage{mathtools}
\usepackage{bbold}
\usepackage{float} 

\DeclarePairedDelimiter\norm{\lvert}{\rvert}      
\DeclarePairedDelimiter\expect{\langle}{\rangle}  

\DeclareSIUnit\atomicmassunit{u}
\DeclareSIUnit\bar{bar}

\usepackage{color}
\usepackage{xcolor}

\newcommand{\Lars}[1]{{\leavevmode \color{red} Lars: #1\newline}}

\begin{document}


\title{Domain formation and structural stabilities in mixed-species Coulomb crystals induced by sympathetically cooled highly charged ions}

\author{L.-A. Rüffert}
\affiliation{Physikalisch-Technische Bundesanstalt, Bundesallee 100, Braunschweig, 38116, Germany}

\author{E. A. Dijck}
\affiliation{Max-Planck-Institut für Kernphysik, Saupfercheckweg~1, 69117~Heidelberg, Germany}

\author{L. Timm}
\affiliation{Institut f\"ur Theoretische Physik, Leibniz Universit\"at Hannover, Appelstraße~2, 30167 Hannover, Germany}

\author{J. R. Crespo López-Urrutia}%
\affiliation{Max-Planck-Institut für Kernphysik, Saupfercheckweg~1, 69117~Heidelberg, Germany}

\author{T. E. Mehlstäubler}
\affiliation{Physikalisch-Technische Bundesanstalt, Bundesallee 100, Braunschweig, 38116, Germany}
\affiliation{Institut für Quantenoptik, Leibniz Universität Hannover, Welfengarten 1, Hannover, 30167, Germany}
\affiliation{Laboratorium für Nano- und Quantenengineering, Leibniz Universität Hannover, Schneiderberg 39, Hannover, 30167, Germany}

\date{\today}

\begin{abstract}
There is a growing interest in high-precision spectroscopy and frequency metrology for fundamental studies using sympathetically cooled highly charged ions (HCIs) embedded in Coulomb crystals of laser-cooled ions.
In order to understand how their strong repulsion affects the crystal structure and dynamics, we study the thermal motion and rearrangement of small mixed linear and homogeneous crystals by both measurements and simulations.
Co-crystallized HCIs form superlattices and divide the crystal into domains, where different reordering rates, melting points and localized phase transitions are observed due to decoupling of motional modes across boundaries.
These results improve our understanding of homogeneous and inhomogeneous ion strings over a wide range of charge-to-mass ratios. This allows us to test our own simulations of the dynamic behavior of ion strings and gives us confidence in their suitability for applications related to quantum simulation as well as computing and the search for new physics. 
\end{abstract}

\maketitle

\section{Introduction}
\label{sec:Introduction}

Trapped ions forming Coulomb crystals \cite{Drewsen1998,Drewsen2000,Drewsen2003} are a subject of intensive research, and find applications in fields such as quantum simulation~\cite{Porras2004, Blatt2008, Friedenauer2008, Roos2012, Barreiro2011, Brooks2022}, high-precision spectroscopy~\cite{Rosenband2008, Yeh2019, Leopold2019, Hausser2024} and tests of fundamental physics~\cite{Safronova2018, Yeh2022, Filzinger2023}.
The achievable control of quantized motional modes together with single-particle resolution in ion Coulomb crystals can be used for the simulation of phase transitions in solid-state physics~\cite{Diedrich1987, Schaetz_2012, Yan2016, Schmidt_Kaler_2013, Ejtemaee2013, Kiethe2021, Pyka2013}.
Tunable confinement strength along the three spatial dimensions controls structural phase transitions, and laser cooling gives access to a wide range of temperatures encompassing plasma, liquid and crystalline phases~\cite{Dubin1996_1, Dubin1996_2, Bedanov1994, Boening2008, Pyka2013, Kiethe2017, Vuletic2015, Kiethe_2018}.
In this way, one can investigate melting in systems far away from the thermodynamic limit, the effects of inhomogeneous particle distances, or the breaking of isotropy by the confining fields, similar to work on dusty plasmas~\cite{Merlino2004, Jaiswal2017, Duca2023}.

Electronic transitions in cold trapped ions are vital as frequency reference for optical clocks, which are finding a growing number of applications in high-precision spectroscopy and tests of fundamental physics~\cite{Huntemann2016, Gerginov2010, Keller2019, Mehlstaeubler2018, Pyka2014, Keller2015PhD}.
The high binding energy of the outer electrons in HCIs allows for extremely narrow electronic transitions also in the extreme ultraviolet range~\cite{Nauta2021} and confers them an innate low sensitivity to external fields~\cite{Kozlov2018}.
HCIs possess many intraconfiguration, forbidden optical transitions of various multiplicities. This makes them attractive candidates for a next generation of optical clocks~\cite{Derevianko2012, Dzuba2012,yudin_magnetic-dipole_2014,Sahoo2018} with a fractional frequency uncertainty of less than $10^{-19}$, and offering transitions with much enlarged sensitivity to a time variation of the fine-structure constant $\alpha$ \cite{berengut_electron-hole_2011,berengut_highly_2012,berengut_optical_2012}, this also including hydrogen-like systems \cite{schiller_hydrogenlike_2007}. Moreover, accurate tests of bound-state quantum electrodynamics can be performed on hydrogen-like, helium-like and lithium-like HCIs with high atomic number~$Z$~\cite{shabaev_stringent_2018, HCI_Safronova_2018, Indelicato2019, Loetzsch2024}.

Lacking allowed optical transitions, sympathetic cooling of highly charged ions (HCIs) was introduced \cite{Schmoeger2015} in order to expand the range of species having forbidden transitions amenable for frequency metrology. In combination with quantum logic \cite{schmidt_spectroscopy_2005}, coherent laser spectroscopy of HCIs, algorithmic cooling and an optical clock based on Ar$^{13+}$ were recently demonstrated \cite{Micke2018,King2021,King2022}.
Sympathetic cooling of HCIs is realized in mixed-species Coulomb crystals. 

Like any other trapped ion species, HCIs are subject to relativistic time dilation shifts caused by their thermal secular motion and the corresponding intrinsic micromotion, which are leading contributions in the uncertainty budget of trapped ion optical clocks~\cite{Keller2015, Keller2019, Yeh2019, King2022}. Additional uncertainties in time dilation may arise from anomalous heating and excess micromotion due to stray fields or RF phase shifts~\cite{Berkeland1998, Brownnutt2015}.


In this paper we theoretically and experimentally investigate the structural stability and dynamics of mixed-species ion Coulomb crystals using their temperature and the trap anisotropy as control parameters.
We investigate the ionic thermal motion, the rate of induced reordering events and their phase transitions.
The paper is organized as follows:
Section~\ref{sec:Theory} introduces the theoretical description of ion dynamics in radio-frequency (RF) traps and discusses existing melting criteria for large crystalline systems, highlighting challenges in applying them to small, inhomogeneous systems.
Section~\ref{sec:Methods} details the numerical simulations and the experimental setup.
As benchmark for mixed-species crystals, Section~\ref{sec:Single-species} analyzes the thermal motion and reordering dynamics of homogeneous ion strings.
We compare our results of molecular dynamics simulations including non-linear effects with the harmonic approximation \cite{Drewsen1998}.
We define a critical rate where thermal reordering dominates over residual gas collisions and discuss the scaling of reordering rates with trapping parameters in single-species crystals.
Section~\ref{sec:Mixed-species} studies the effects of HCIs on the dynamics of mixed-species crystals in different temperature regimes, revealing localized melting behavior due to HCIs.
Finally, Section~\ref{sec:Conclusion} concludes our results.

\section{Theoretical background}
\label{sec:Theory}

\subsection{Ion Coulomb crystals}
\label{subsec:ICC}

We consider a system composed of $N$ ions with positions $\vec r_i=\left(x_i,y_i,z_i\right)$, masses~$m_i$ and (positive) charges~$Q_i$ ($i=1,\dots,N$) confined in a linear RF trap.
In the ponderomotive approximation, the static and oscillating electric fields of the trap generate a confining potential which, to lowest order of a Magnus expansion, is described by a time-independent quadratic form in all three dimensions~\cite{Paul1990}.
In addition, the ions repel each other due to the Coulomb force which yields a total potential
\begin{align} \label{eq:potential}
    \mathcal{V} = \sum_i^N \frac{1}{2} m_i \left( \omega_{x,i}^2 x_i^2 + \omega_{y,i}^2 y_i^2 + \omega_{z,i}^2 z_i^2 \right) + \sum_{i<j}^N \frac{Q_i Q_j}{4\pi \epsilon_0 d_{ij}} \text{,}
\end{align}
where $\epsilon_0$~is the vacuum permittivity and $d_{ij} = \norm{\vec{r}_i - \vec{r}_j}$ is the distance between ions~$i$ and $j$.
The secular frequencies $\omega_{x,i}$, $\omega_{y,i}$ and $\omega_{z,i}$ carry a subscript~$i$ as the confinement strengths experienced by different ions stored in the same trap depend on their charge and mass.
We define the axial trap direction with only static electric potential as $z$-axis, so that the secular frequencies are given by
\begin{align}
    \begin{split}
        \label{eq:omegas}
        \omega_{z,i}^2 &= \frac{Q_i}{m_i}u_\text{DC} \\
        \frac{\omega_{x/y,i}^2}{\omega_{z,i}^2} &= \frac{1}{2}\frac{Q_i}{m_i} \frac{u_\text{RF}^2}{u_\text{DC} \Omega_\text{RF}^2} - \frac{1}{2} \mp c_{xy},
    \end{split}
\end{align}
where the negative sign of the $\mp$ symbol is used to calculate $\omega^2_{x,i}/\omega^2_{z,i}$, while the positive sign corresponds to $\omega^2_{y,i}/\omega^2_{z,i}$. Here, $u_\text{DC}$ ($u_\text{RF}$)~denotes the magnitude of the static (oscillating) electric field gradient in the trap center, $\Omega_\text{RF}$~is the angular frequency of the oscillating component and $c_{xy}$~quantifies the difference between the two radial secular frequencies.
Throughout the paper we focus on the case $\omega_{y,i} > \omega_{x,i} > \omega_{z,i}$ so that the weakest radial confinement which is of main interest for the melting dynamics is exerted along the $x$-axis.
The resulting confining force along the $z$-axis is proportional only to the charge of the ion, whereas the oscillating field in the radial direction produces a non-linear dependence of the restoring force on both charge and mass of the ion.
As a result, HCIs are pushed towards the trap center more strongly.

At vanishing temperature, the ions settle at equilibrium positions~$\left\{\vec{r}_i^{\,0}\right\}$, forming a crystal with a shape dictated by the competition between the Coulomb force pushing ions apart and the tunable trap potential restoring ions to the trap center.
In this paper we focus on the simplest case, i.e.\ the linear phase which is entered when the trapping frequencies are brought into a regime $\omega_{x,i} \gg \omega_{z,i}$ and the ions equilibrate in a chain with $x_i^0 = y_i^0 = 0$.

For generality we use dimensionless quantities throughout the paper by defining
\begin{align} \label{eq:rescaled_length}
    \ell = \sqrt[3]{\frac{e^2}{4 \pi \varepsilon_0 m' \omega_{z}'^2}} \quad \text{and} \quad \tau = \frac{m'\omega_z'^2\ell^2}{k_B}
\end{align}
as unit of length and temperature, respectively.
Here, $e$ is the electron charge, $m'$ is the mass of the ion species in the system chosen as a reference, $\omega_z'$ ($\omega_x'$) is the axial (radial) secular trapping frequency experienced by these ions and $k_B$ is the Boltzmann constant.
We will quantify the axial--radial anisotropy of the trap by
\begin{align}
    \alpha = \frac{\omega_x'^2}{\omega_z'^2} \text{.}
\end{align}

\subsection{Crystal melting criteria}
\label{subsec:Melting_Criteria}

Various criteria have been used in the literature to define the melting transition of crystalline systems.
One of the longest-established is the Lindemann criterion, which is based on the vibrational amplitude of atoms within a crystal lattice~\cite{Lindemann1910}.
It postulates that melting occurs when the mean amplitude of thermal vibrations of atoms reaches a critical fraction of the average inter-atomic spacing.
This melting criterion can be expressed as
\begin{equation} \label{eq:Lindemann}
    \frac{\sqrt{\langle \vec r_i^{~2}\rangle-\langle \vec r_i\rangle^2}}{d} = C_L \text{,}
\end{equation}
where $d$ is the average nearest-neighbor distance and $C_L$ is the Lindemann constant, typically set between $0.1$ to $0.2$ depending on the crystal structure, the specific constituents and interaction potential~\cite{Stillinger1980_Lindemann, Agrawal1995_Thermodynamic, Agrawal1995_Thermodynamic_2, Chakravarty2002_Integral, Batsanov2009_Melting, Navarro2011}.

A second way to characterize the thermodynamic state of one-component plasmas is based on the spatial correlation between particles.
This is quantified by a dimensionless coupling parameter~$\Gamma$, defined as the ratio of the interaction energy between neighboring particles and the thermal energy $k_B T$.
In the case of Coulomb-interacting charges~$Q$, it is given by
\begin{equation} \label{eq:Gamma}
    \Gamma = \frac{Q^2}{4\pi\epsilon_0 d_S k_B T} \text{,}
\end{equation}
where $d_S$ denotes the Wigner--Seitz radius
\begin{equation}
    d_S = \sqrt[3]{\frac{3}{4 \pi n}}
\end{equation}
for mean particle density~$n$.
Most classical plasmas are weakly coupled, characterized by~$\Gamma < 1$.
Due to efficient laser cooling methods, ion Coulomb crystals can reach values of~$\Gamma \gg 1$ with strongly correlated particle motion that in nature are only found in extreme environments such as the interior of heavy planets or the outer crust of neutron stars~\cite{Ichimaru1982}.
By looking at the functional relation between the Helmholtz free energy and the coupling parameter~$\Gamma$ for the liquid and solid phase of one-component plasmas, their melting transition was found to be around $\Gamma_m = 168\pm4$~\cite{Pollock1973}.

Both of these criteria were derived for large, isotropic systems of a single particle species.
Macroscopic systems undergo phase transitions based on the collective behavior of a vast number of particles.
In contrast, in microscopic systems, they are governed by the dynamics of a small number of particles.
Here we consider linear ion Coulomb crystals of up to a few dozen ions.
This case differs in several key aspects from the aforementioned larger systems: different trapping frequencies along the three spatial directions, necessary for the production of linear ion strings, break the isotropy of the particle dynamics.
The considered harmonic confinement causes boundary effects, i.e., inhomogeneity in the particle separations and thermal vibrations.
Such effects are considerably enhanced by the presence of HCIs, as demonstrated in this paper.
Hence, the concept of phase transition becomes more nuanced, requiring different theoretical measures and interpretations.
Finding alternative ways to characterize the melting of linear, single-species as well as mixed-species Coulomb crystals with a small number of ions will be the focus of this work.

\section{Numerical and experimental methods}
\label{sec:Methods}

We perform molecular dynamics (MD) simulations \cite{Nigmatullin_2022, Okada_2014, Ozawa_motional_2019}, as well as experiments. For the former, we employ the Langevin formalism to describe the dynamics of ions in RF traps at finite temperature.
The equations of motion incorporate stochastic forces reflective of Brownian motion and a drag force to model laser cooling.
All simulations are initialized at $T=0$ to find the equilibrium positions of the ions before raising the temperatures allowing for thermalization of the ensemble.
Time steps are carefully chosen to prevent numerical errors and ensure accurate time-averaging of the ion dynamics.
At low temperatures, the motion of the ions approximates that of coupled harmonic oscillators, allowing comparisons between analytic and simulated results.
More details on the simulations can be found in Appendix~\ref{sec:simulations_methods}.


\begin{figure}
  \includegraphics{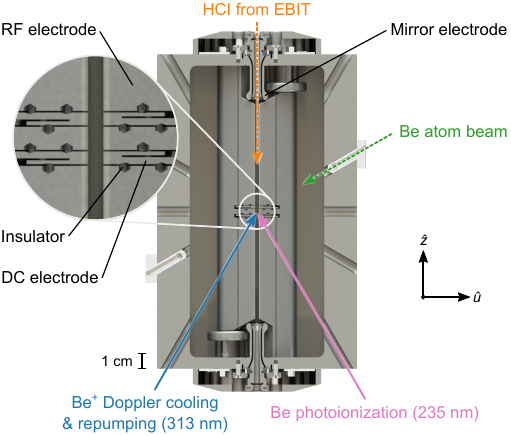}
  \caption{%
Top-down section view of the CryPTEx-SC Paul trap matching the perspective of the imaging optics, with coordinate system marked.
Ports in the superconducting RF resonator body provide optical access.
HCIs are initially confined between mirror electrodes at either end of the RF blades until capture within a pre-loaded Be$^+$ ion crystal trapped between the DC electrodes.
}
  \label{fig:experiment_setup}
\end{figure}

To benchmark the simulations, we compare them with the dynamics observed in ions trapped in a linear Paul trap.
The cryogenic ion trap of the CryPTEx-SC experiment (see Fig.~\ref{fig:experiment_setup}) is of quasi-monolithic design detailed elsewhere~\cite{Stark2021, Dijck2023}, operating at $\Omega_\text{RF}/2\pi = \SI{35}{\mega\hertz}$.
Singly charged Be$^+$ ions are loaded by photo-ionization from a $^9$Be atom beam, while HCIs are extracted from an electron beam ion trap (EBIT) and transferred to the Paul trap using ion optics and a set of pulsed drift tubes for deceleration.
Doppler cooling is performed on the Be$^+$ ions using laser light at \SI{313}{\nano\meter} under an angle of \SI{30}{\degree} to the trap axis to cool all motional modes of the ion crystal.
Ion fluorescence is captured using cryogenic imaging optics focusing light onto an EMCCD camera (electron-multiplying charge-coupled device).
Further details of the experimental setup are provided in Appendix~\ref{sec:appendix_experimental_setup}.

\section{Single-species linear ion crystals}
\label{sec:Single-species}

\subsection{Simulation of hopping dynamics}
\label{subsec:Single-species_simulations}

\begin{figure}
  \includegraphics[width=\columnwidth]{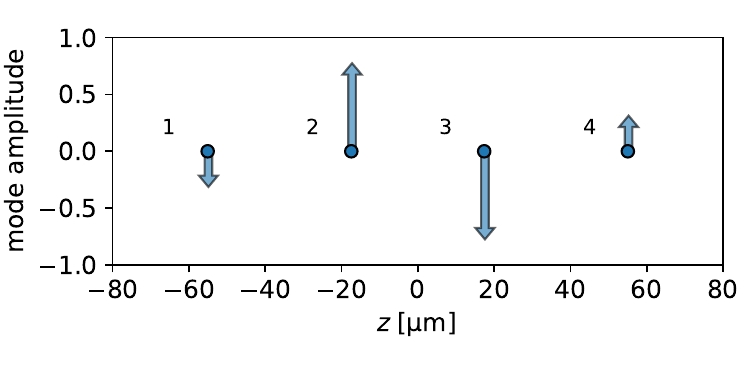}
  \caption{%
Numerically found equilibrium positions of a four Be$^+$ ion crystal at $\omega_z'/2\pi = \SI{85.6}{\kilo\hertz}$ along the $z$-axis (circles).
Additionally, the ion motion in the out-of-phase 'zigzag' mode is shown along the most weakly confined radial direction (indicated by arrows).
}
  \label{fig:4_Be_zigzag_mode}
\end{figure}

To illustrate our methods and discuss the relevant dynamics, we first employ molecular dynamics simulations of a linear string of four ions of a single species.
We here consider singly charged Be$^+$ ions ($m' = \SI{9.012}{\atomicmassunit}$) at an axial secular frequency of $\omega_z'/2\pi = \SI{85.6}{\kilo\hertz}$.
We consider $\omega_x,\omega_y \gg \omega_z$ so that the ions form a linear ion chain along the $z$-axis of the trap, with $\omega_x'/2\pi = \SI{199.4}{\kilo\hertz}$ and $\omega_y'/2\pi = \SI{271.4}{\kilo\hertz}$. These frequencies match experimental data; we discuss their scaling with trapping parameters later.
Throughout the paper we set the damping rate to $m' \eta = \SI{1e-20}{\kilogram\per\second}$ (see Appendix~\ref{sec:simulations_methods}).
As we consider ions in thermal equilibrium and neglect collisions with residual gas particles, our results are not sensitive to the selected rate.
Figure~\ref{fig:4_Be_zigzag_mode} shows the simulated four-ion crystal at $T=0$ with all ions at their equilibrium positions.

We quantify the three-dimensional thermal motion of the ions at $T>0$ using the normalized average root-mean-square (RMS) displacement as
\begin{align}
  s_\ell = \frac{1}{N} \sum_i\frac{\sqrt{ \langle{ \left( \vec{r}_i - \vec{r}_i^{\,0} \right)^2 \rangle} }}{\ell} \text{,}
\label{eq:sigma}
\end{align} 
where $\ell$~represents the characteristic length scale of the system defined by Eq.~\eqref{eq:rescaled_length} and $\vec{r}_i^{\,0}$ denotes the equilibrium ($T=0$) position of the particles.
We note that we always keep the initial equilibrium position of each ion as reference point for the calculation of~$s_\ell$, which becomes relevant at temperatures at which the ions can change their $z$-ordering.

\begin{figure}
\includegraphics[width=\columnwidth]{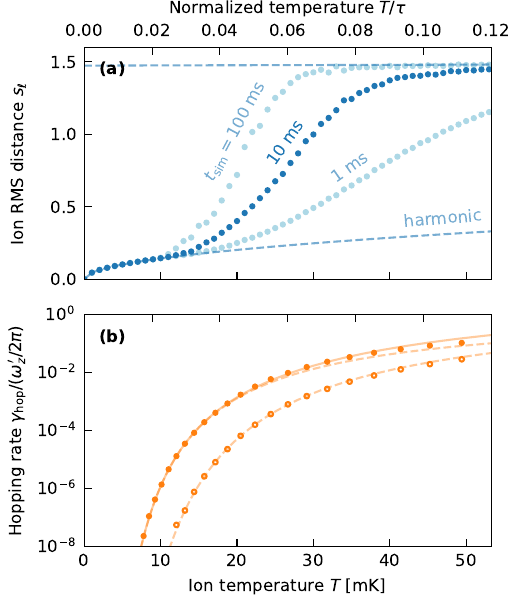}
\caption{%
Molecular dynamics simulation of a four Be$^+$ ion string. (a)~Average normalized RMS ion motion extent~$s_\ell$ (see text) as a function of the system temperature~$T$ for a simulated time~$t_\text{sim}$ of \SIrange{1}{100}{\milli\second} (dots). Lower dashed curve: ion motion extent computed in second order harmonic approximation. Upper dashed curve: saturation value assuming continuous reordering of the ions.
(b)~Normalized hopping rate for the inner pair (ions 2--3, filled dots) and outer pairs (ions 1--2, 3--4, open dots) separately.
Dashed curves show low-temperature fits with Eq.~\eqref{eq:exp_factor}. Solid curve: sum total hopping rate (at low~$T$ dominated by hopping of the inner pair).
Simulated secular frequencies are $\omega_{x,y,z}'/2\pi =  \{ 2.33, 3.17, 1.0 \} \times \SI{85.6}{\kilo\hertz}$. 
}
\label{fig:Be_Lindemann}
\end{figure}

In Fig.~\ref{fig:Be_Lindemann}(a) we show the numerically calculated~$s_\ell$ for three different simulation lengths~$t_\text{sim}$ (averaged over repeated runs) as function of system temperature~$T$ and compare the results with the harmonic approximation that scales exactly as~$\sqrt{T}$ (see Appendix~\ref{sec:analytic_msd}).
Depending on the temperature, we can distinguish three different regimes.
At low temperatures, the numerical result which includes non-linear effects agrees well with the harmonic approximation.
In this regime, the particles oscillate around their initial equilibrium position with a temperature-dependent mean amplitude.
Evaluating Eq.~\eqref{eq:sigma} for a simulation time of $t_\text{sim} = \SI{10}{\milli\second}$, the numerical result starts deviating from the $\sqrt{T}$~scaling at a normalized temperature of $T/\tau \approx 0.03$.
At this combination of $t_\text{sim}$ and $T$, the probability for ions to alter their $z$-ordering during the simulated time, which we will refer to as `hopping', becomes noticeable.
Even if only a single hopping event takes place in a simulation run, it abruptly increases the mean oscillation amplitude~$s_\ell$.
As hopping events become more frequent with temperature in this second regime, $s_\ell$~increases until saturating at $T/\tau \approx 0.12$.
The maximum value reached in the third regime is indicated by the upper dashed line in Fig.~\ref{fig:Be_Lindemann}(a), calculated by evaluating Eq.~\eqref{eq:sigma} without distinguishing ion trajectories $\vec{r}_i$, i.e.\ associating each ion with equal probability with each of the equilibrium positions.
The temperature at which~$s_\ell$ starts deviating from the harmonic approximation depends on the product of the hopping rate and the chosen simulation time~$t_\text{sim}$ as illustrated by the additional data in Fig.~\ref{fig:Be_Lindemann}(a) (grey points). 
Extracting a well-defined melting temperature of the crystal structure from~$s_\ell$ is therefore not possible.
Nonetheless, the qualitative behavior is consistent across all values of~$t_\text{sim}$.

Associated with the behavior of the size of particle fluctuations, we show the rate of hopping events~$\gamma_\text{hop}$ per ion pair as function of temperature in Fig.~\ref{fig:Be_Lindemann}(b).
We calculate~$\gamma_\text{hop}$ by counting instances at which the $z$-ordering of ion positions changes in the simulations.
In contrast to~$s_\ell$, the hopping rate is independent of simulated time~$t_\text{sim}$.
It continuously increases with $T$ and is never exactly zero at $T>0$ due to the exponential tail of the thermal energy distribution of the ions, though in practice bound by the inverse of the simulated time. Its dependence on $T$ agrees well with the exponential scaling of a thermal process with an energy barrier, as in the Arrhenius equation
\begin{equation} \label{eq:exp_factor}
  \gamma_\text{hop}(T) = A \exp \left( -\Delta E_N(\alpha)/k_B T \right) \text{,}
\end{equation}
where the fit parameters are limiting rate~$A$ (on the order of $\omega_z'/2\pi$) and energy $\Delta E_N(\alpha)$ needed to hop between lattice sites in a crystal with $N$~ions.
We will discuss the $\alpha$-dependence of the energy barrier in further detail in Section~\ref{subsec:Scaling}.
The results of fitting Eq.~\eqref{eq:exp_factor} to our simulated data are depicted as dashed lines in Fig.~\ref{fig:Be_Lindemann}(b).


A closer look at the hopping rate reveals effects of the inhomogeneity in the system:
when evaluating it for individual neighboring ion pairs, a difference is observed between the rate at which the two innermost ions (ion 2 and 3 in Fig.~\ref{fig:4_Be_zigzag_mode}) switch positions and that of the outer pairs (ions 1--2 and 3--4).
This is explained by the difference in energy required to trigger the hopping of the ion pairs, with the fits in Fig.~\ref{fig:Be_Lindemann}(b) indicating an energy barrier of $\Delta E = \SI{12.1}{\micro\electronvolt}$ for the inner pair and \SI{18.9}{\micro\electronvolt} for the outer pairs in this example.
A second observation is that the simulated hopping rates match Eq.~\eqref{eq:exp_factor} within statistical uncertainty at low temperatures, but start deviating above $T/\tau\approx 0.06$.
We note that our definition of a hopping rate ($z$-reordering) ceases to be meaningful at high temperatures where a crystalline structure can no longer be discerned.

At low temperatures, the total hopping rate and the fluctuation amplitude~$s_\ell$ are dominated by hopping of the innermost ions.
Signatures of this behavior can also be seen in the motional mode vector of the radial out-of-phase oscillations depicted in Fig.~\ref{fig:4_Be_zigzag_mode}, often referred to as the zigzag mode~\cite{Kiethe2021,Fishman2008}.
This mode has the lowest frequency of the radial set of modes and has therefore the largest thermal excitation amplitude.
It has the highest projection onto the innermost ions, thus increasing the hopping probability of these ions in particular.


\subsection{Comparison with experimental data}
\label{subsec:Single-species_Experimental_comparison}

\begin{figure}
\includegraphics[width=\columnwidth]{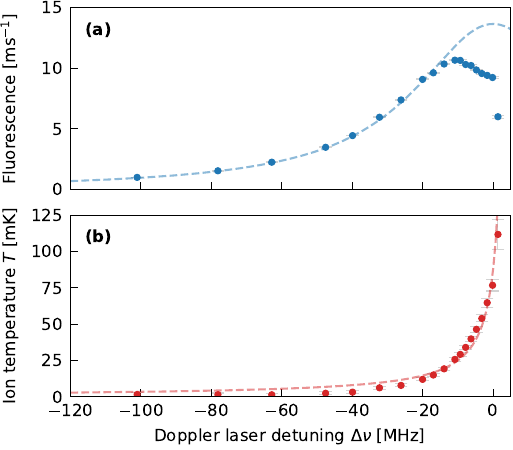}
\caption{%
(a)~Measured fluorescence rate per ion for a four Be$^+$ ion string as function of Doppler laser detuning~$\Delta\nu$.
At $\Delta\nu=0$, the ions are no longer cooled and fluorescence disappears.
The fit yields a saturation parameter of $S \approx 8.3$.
(b)~Ion temperature determined from EMCCD images using spatial thermometry, approximately following a $1 / \norm{\Delta\nu}$ curve (dashed line).
}
\label{fig:exp_doppler_temp}
\end{figure}

We now compare the molecular dynamics simulations of four Be$^+$ ions with experimental data in the form of spatially-resolved EMCCD ion fluorescence images.
One axis of the image plane is aligned with the $z$-direction of the ion trap (within \SI{2}{\degree}) and the other axis, which we will denote as~$u$, corresponds to a projection of both the $x$ and $y$ radial directions under~\SI{45}{\degree} (see Fig.~\ref{fig:experiment_setup}).
Changing the detuning of the cooling laser while keeping the repumper laser frequency fixed allows to reproducibly set the temperature of the ion crystal over a wide range.
The average fluorescence rate at different detunings~$\Delta\nu$ of the Doppler laser is shown in Fig.~\ref{fig:exp_doppler_temp}(a).
Away from the atomic resonance, the fluorescence rate follows a saturation-broadened Lorentzian spectrum, while thermal broadening distorts the spectrum at small detunings.

To determine the ion temperature at a particular laser detuning, we use spatial thermometry~\cite{Knuenz2012,Rajagopal2016,Dijck2023} to extract it from the EMCCD images.
Here we determine the (projected) spatial extents~$\sigma_{u,i}$, $\sigma_{z,i}$ of all ions by simultaneously fitting the spatial fluorescence intensity profile of each ion with a two-dimensional Gaussian function in addition to a constant background term.
We choose fairly low secular frequencies to increase the size of thermal motion on the camera images.
The observed spot sizes are a convolution of the projected ion spatial distributions with the point spread function (PSF) of the imaging system.
We assume the PSF to also be Gaussian so that
\begin{equation} \label{eq:spatial_extents_psf}
    \sigma_{u,i} = \sqrt{(\sigma^\text{PSF}_u)^2 + (\sigma^\text{th}_{u,i})^2} \text{, } \sigma_{z,i} = \sqrt{(\sigma^\text{PSF}_z)^2 + (\sigma^\text{th}_{z,i})^2} \text{,}
\end{equation}
where $\sigma^\text{th}_{u,i}$ and $\sigma^\text{th}_{z,i}$ denote the (projected) spatial extents due to thermal ion motion.
The former is related to motion along the principal trap axes by
\begin{equation}
    \sigma^\text{th}_{u,i} = \sqrt{\frac{(\sigma^\text{th}_{x,i})^2 + (\sigma^\text{th}_{y,i})^2}{2}} \text{.}
\end{equation}
The thermal and PSF contributions to the observed spot sizes can be distinguished, as the former depends on temperature~$T$ and which ion~$i$ is considered, while the PSF contribution is taken to be constant.
The result remains well-defined even if (moderate) ion hopping takes place and the ion spatial distributions start overlapping. 
Note that if no hopping takes place and the time-averaged ion spatial distributions were exactly Gaussian, $\sigma^\text{th}_i$ summed in quadrature over the three spatial directions and averaged over all ions would equal $\ell s_\ell$.

\begin{figure*}[t]
  \includegraphics[width=\textwidth]{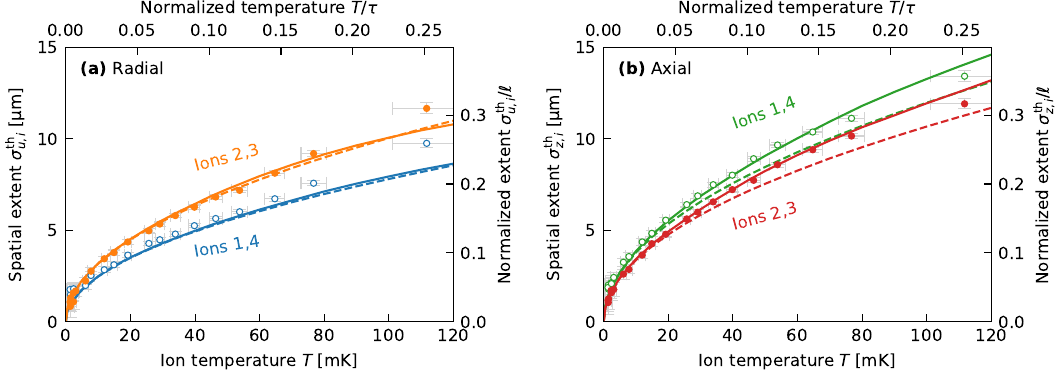}
  \caption{%
Benchmark of the simulation by comparing the simulated spatial thermal extent (solid lines) to experimental spot sizes (points) as well as the value predicted by the harmonic approximation (dashed lines). We compare the radial and axial extents of the outer (1, 4) and inner ions (2, 3) of the four Be$^+$ ion crystal in (a) and (b), respectively. The secular frequencies are $\omega_{x,y,z}'/2\pi = \{ 2.33, 3.17, 1.0 \} \times \SI{85.6}{\kilo\hertz}$.
Ion temperature is determined by matching fitted $u$, $z$ extents from EMCCD images to simulation results.
}
\label{fig:4be_spatial_extents}
\end{figure*}

To test the validity of the simulations, we compare in Fig.~\ref{fig:4be_spatial_extents} their results with the spot sizes in radial and axial direction found in the experiment using the same fitting procedure.
The spot size of the inner two ions is larger than that of the two outer ones due to boundary effects; for the axial spatial extent, this is reversed.
In addition, we depict the harmonic approximation (dashed lines, see Appendix~\ref{sec:analytic_msd}), which converges with the simulation results at low temperature.
Deviation due to non-linear effects of the Coulomb force is here particularly seen in the axial sizes (Fig.~\ref{fig:4be_spatial_extents}(b)), where the non-linear contribution reaches up to~\SI{10}{\percent} within the considered temperature range.

Although the analytic harmonic approximation~\eqref{eq:harmonic_msd} can be used for spatial thermometry and yields accurate results at temperatures close to the Doppler temperature~\cite{Knuenz2012}, we here apply the same 2D fitting procedure to simulated data to take additional ion dynamics into account.
The relative spatial extents observed in experiments below~\SI{60}{\milli\kelvin} are consistent with the full simulation, and we attribute some deviation in the spot sizes observed at higher temperatures to limited precision of the laser detuning control, which could cause some fluctuation in the ion temperature when the dependence on detuning becomes very steep.
Using the analytic harmonic approximation for spatial thermometry may lead to significant errors in the extracted value of~$T$ at higher temperatures. 
A deeper analysis of the non-linear corrections to the Gaussian distributions of the ions is out of the scope of this paper.
With the described technique, we determine the ion temperature, shown in Fig.~\ref{fig:exp_doppler_temp}(b) to be roughly proportional to the inverse of the laser detuning~$\norm{\Delta\nu}$.

\begin{figure*}[t]
  \includegraphics[width=\textwidth]{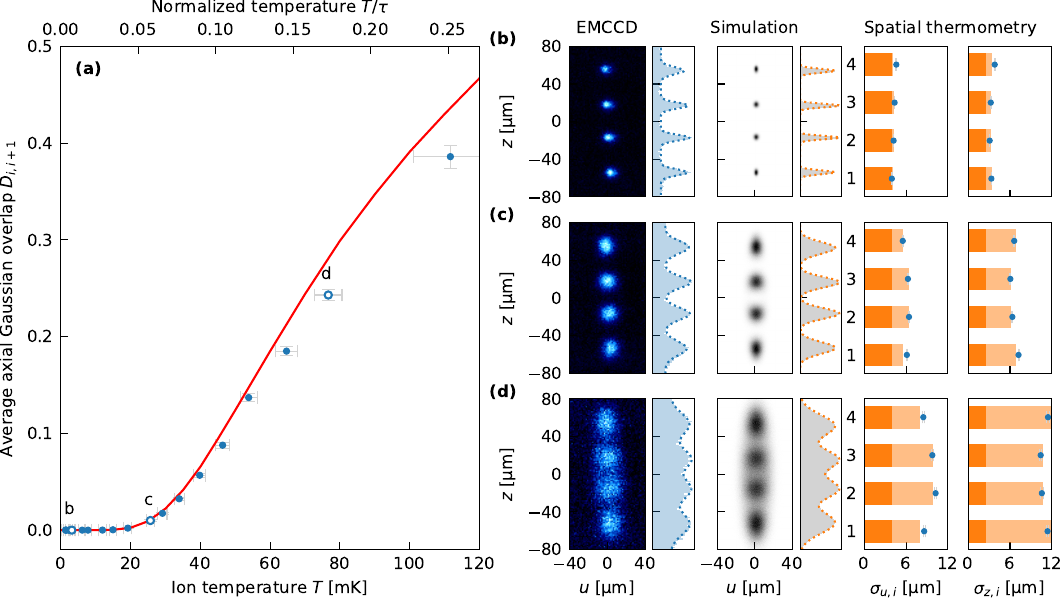}
  \caption{%
Average Gaussian overlap $D_{i,i+1}$ of neighboring ion axial distributions from simulation (solid line)
and experiment (points) for a four Be$^+$ ion string with $\omega_{x,y,z}'/2\pi = \{ 2.33, 3.17, 1.0 \} \times \SI{85.6}{\kilo\hertz}$.
Insets (b), (c) and (d) show data at three marked temperatures in greater detail: fitted \SI{20}{\second} exposure time EMCCD image, simulated data and spatial thermometry fit to determine~$T$ where the dark part of bars indicates fitted PSF contribution. See main text for details.
}
  \label{fig:exp_4be}
\end{figure*}

Neither of the simulated quantities presented in Fig.~\ref{fig:Be_Lindemann} to illustrate the reordering dynamics are directly observable in experiment, as we cannot distinguish ions of the same species.
Therefore, we instead compare the overlap of axial Gaussian distributions here.
We define the dimensionless overlap~$D_{ij}$ (Bhattacharyya coefficient) between ions~$i$ and $j$ as the (one-dimensional) overlap integral of fitted axial spatial distributions~$P_i$ and $P_j$:
\begin{align}
    D_{ij} &= \int{\sqrt{P_i(z) P_j(z)} dz} \\ &= \sqrt{\frac{2 \, \sigma^\text{th}_{z,i} \, \sigma^\text{th}_{z,j}}{(\sigma^\text{th}_{z,i})^2 + (\sigma^\text{th}_{z,j})^2}} \exp \left[-\frac{(z_i - z_j)^2}{4 [(\sigma^\text{th}_{z,i})^2 + (\sigma^\text{th}_{z,j})^2]}\right] \text{,} \nonumber
\end{align}
where $z_i$ and $z_j$ are the axial mean positions of the ions fitted from the experimental data.
Although this quantity does not directly give the probability for ion hopping, as this would require information on the correlation of ion motion unavailable from the EMCCD data, it serves as a practical proxy for assessing the probability of finding two ions at the same $z$~coordinate, and is therefore related to the hopping rate.


Figure~\ref{fig:exp_4be}(a) compares the overlap~$D_{i,i+1}$ of axial spatial extents between experiment and simulation, averaged over the three ion pairs as function of ion temperature~$T$.
Uncertainties are ($\chi^2$-corrected) standard errors from the fit.
Representing the data in this way shows that the axial overlap of ion motion is negligible below about $T/\tau \approx 0.04$, and then starts increasing, concurrent with the hopping rate as shown in Fig.~\ref{fig:Be_Lindemann}(b).
Fitting of data and simulation is shown in more detail for selected data points in Figs.~\ref{fig:exp_4be}(b)--(d).
EMCCD images and simulated data histograms are fitted with a two-dimensional Gaussian function for each ion position (dotted lines in the projections next to the two-dimensional plots); the EMCCD fit also includes a constant background term.
The rightmost panels show the spatial thermometry fits used to determine the temperature at each detuning: the fit model based on interpolated simulation results is represented by the colored bars and the data points are the radial and axial~$\sigma$~widths of the two-dimensional Gaussian functions fitted to the EMCCD data.
Free parameters are the ion crystal temperature~$T$ at each detuning and two global parameters for the PSF contribution in radial and axial directions, see Eq.~\eqref{eq:spatial_extents_psf}, found to be $\sigma^\text{PSF}_u = \SI{3.9}{\micro\meter}$ and $\sigma^\text{PSF}_z = \SI{2.6}{\micro\meter}$, respectively.
We note that the calculated Gaussian overlap is not independent of the temperature assigned to each data point, as the fitted spatial extents are used in both.

In conclusion, we find that the simulation including non-linear effects qualitatively matches the experimental data well.
Some systematic offset between simulation and experiment can be attributed to effects from optical aberrations not accounted for in the simplified Gaussian PSF model, a possible slight tilt of the crystal out of the focal plane and the illumination of the ions by laser light not being perfectly uniform.

\subsection{Scaling of hopping rate with ion number and trap confinement}
\label{subsec:Scaling}

Our result of a continuously increasing ion hopping rate (see Fig.~\ref{fig:Be_Lindemann}(b)) does not reveal a distinct threshold indicating the onset of melting in a linear ion crystal.
Consequently, how to define the melting temperature is not readily apparent.
A rigorous definition is only possible in the thermodynamic limit, taking the number of ions~$N$ to infinity while the axial trapping frequency~$\omega_z'$ approaches zero, such that the ion spacing~$d$ remains constant.
In this limit, a first-order melting transition can be discerned from the calculation of the free energy of the system~\cite{DeWitt_1982}.
In finite, experimentally accessible systems, a crossover window is the remainder of this transition.
Defining a melting temperature based on the magnitude of ion oscillations~$s_\ell(T)$ is problematic due to its dependency on the simulation length.
One proposed method is based on the variance of block-averaged inter-particle distances, however, this similarly requires choosing an appropriate time length for the individual blocks~\cite{Boening2008}.
Several advanced strategies for the computation of melting temperatures have been developed in the context of dusty plasmas~\cite{thomsen_resolving_2015, melzer_phase_2012, melzer_instantaneous_2012, melzer_analyzing_2013}.

Instead of identifying a melting temperature in the strict sense, we define a critical threshold temperature~$T_c$ as the point where the thermal hopping rate becomes large compared to hopping events caused by collisions with residual gas particles.
Typical room-temperature ion trap experiments operate at a pressure between \SIrange[print-unity-mantissa=false]{e-10}{e-11}{\milli\bar}.
The rate of external collisions is usually $\lesssim \SI{0.01}{\per\second}$ per ion at an axial secular frequency of $\omega_z'/2\pi = \SI{100}{\kilo\hertz}$ and radial secular frequencies on the order of $\simeq\SI{1}{\mega\hertz}$~\cite{Aikyo2020}.
We therefore define a generalized critical hopping rate as
\begin{equation}
    \gamma_c \approx 10^{-7} \times \omega_z'/2\pi \text{.}
\end{equation}
Although somewhat arbitrary, this enables us to quantify the effects of ion number, trapping parameters and presence of HCIs on the reordering dynamics of the crystals.
A discussion comparing our critical temperature definition with the melting criteria introduced in Section~\ref{subsec:Melting_Criteria} is presented in Appendix~\ref{sec:Lindemann_Gamma_discussion}.

\begin{figure}
  \includegraphics[width=\columnwidth]{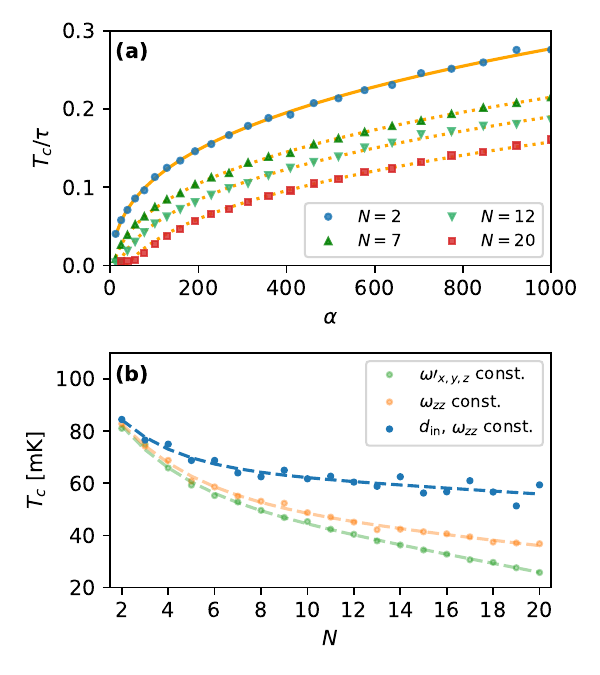}
\caption{(a)~$T_c/\tau$ as function of~$\alpha$ for $N=\{2,7,12\}$, where the $N=2$ data is fitted with a function $\propto (\sqrt[3]{\alpha}-1)$ (solid line) (b)~(absolute) critical temperature~$T_c$ as a function of~$N$, illustrating convergence when choosing $\alpha(N)$ and $\omega_z'(N)$ such that the spacing $d_\text{in}$ of the innermost ion pair and the zigzag mode frequency $\omega^{zz}$ remain constant.
}
\label{fig:PureStringsHeatmap}
\end{figure}


Let us now discuss what influences the occurrence of hopping, first concentrating on trap confinement.
The Coulomb repulsion prevents ions from approaching each other, so a change in their $z$-ordering happens by buckling out into one of the radial dimensions.
This requires a minimum energy given by the (lowest) radial secular frequency.
Consequently, hopping starts to occur when the thermal energy of the ions is comparable to the increase in potential energy during the hop.
This threshold can be estimated by taking the difference~$\Delta E$ between the equilibrium configuration and the minimal-energy configuration at which the considered ions have the same $z$-coordinate.
In the case~$N=2$ an analytical derivation yields
\begin{align} \label{eq:energy_barrier}
  \frac{\Delta E_{N=2}(\alpha)}{m'\omega_z'^2\ell^2} = \frac{3\sqrt[3]{4}}{4} \left( \sqrt[3]{\alpha} - 1 \right) \text{,}
\end{align}
which vanishes for~$\alpha=1$ as this indicates the case of rotational symmetry in the $z$--$x$ plane~\cite{Duca2023}.
This analytically derived expression can be compared to values of~$\Delta E$ estimated from simulated hopping rates using Eq.~\eqref{eq:exp_factor} and found to agree within about~\SI{5}{\percent}.
The energy barrier can alternatively be parametrized in terms of the zigzag mode frequency~$\omega^{zz}$ since $\alpha = (\omega^{zz}/{\omega_z'})^2 + 1$ for a two-ion crystal.
For larger ion numbers~$N$, numerical results show that the energy barrier additionally depends on the specific pair of ions considered.
Figure~\ref{fig:PureStringsHeatmap}(a) shows the scaling of the critical temperature~$T_c$ with the trap anisotropy parameter~$\alpha$.
Numerical results for $N=2$ agree very well with the $\sqrt[3]{\alpha}$ scaling of the energy barrier expected from Eq.~\eqref{eq:energy_barrier}.
We find the thermal energy of the ions at the critical temperature to equal about $0.08 \times \Delta E_{N=2}(\alpha)$.
Note that in the limit $\alpha\rightarrow\infty$ the critical temperature diverges as the system becomes one-dimensional.
Numerical results for $N>2$ suggest that the scaling with~$\alpha$ of the energy barrier retains a similar form as Eq.~\eqref{eq:energy_barrier} with only a change to the $-1$~offset within parentheses.

The phase-space probability in the non-equilibrium steady state follows the Maxwell--Boltzmann distribution.
As already seen in Fig.~\ref{fig:Be_Lindemann}(b), hopping rates at low temperature follow Arrhenius equation~\eqref{eq:exp_factor} with a pre-exponential factor~$A$ of order $\omega_z'/2\pi$.
Thus, the energy barrier for hopping (of the innermost ions) can be related to the critical temperature as
\begin{equation}
\label{eq:energy_barrier_estimation}
\Delta E \approx -k_B T_c \ln{10^{-7}} \approx 16.1 \times k_B T_c \text{,}
\end{equation}
enabling the estimation of hopping rates from Fig.~\ref{fig:PureStringsHeatmap}(a).
If a more precise extrapolation is required, additional simulations can be performed to extract the relevant energy barrier~$\Delta E_N$ and pre-exponential factor~$A$.
We find that simulating the ion dynamics at temperatures with the hopping rate between $\gamma_\text{hop} \approx 10^{-4} \, \omega_z'/2\pi$ and $10^{-3} \, \omega_z'/2\pi$ suffices for a precise extrapolation of the hopping rate to lower temperatures using Eq.~\eqref{eq:exp_factor} without requiring too long simulations.

With increasing number of ions~$N$, the spacing between ions as well as the frequency~$\omega^{zz}$ of the zigzag mode is reduced, resulting in a lower energy barrier, and therefore lower critical temperature~$T_c$.
This is illustrated by the lower trace in Fig.~\ref{fig:PureStringsHeatmap}(b).
When keeping $\omega^{zz}$ constant by tuning the radial trapping frequency (middle trace), $T_c$ drops less than in the first case where all trapping frequencies were kept constant.
Nonetheless, in both cases $T_c$ eventually drops to zero when increasing the number of ions further.
When additionally keeping the distance of the inner ions $d_\text{in}$ constant by tuning the axial frequency (upper trace), we find that $T_c$ shows an even shallower drop, potentially reaching a finite value for $N \rightarrow \infty$.
A thorough analysis of this limit is beyond the scope of this paper.

For any single-species linear ion Coulomb crystal, the time-averaged dynamics depend only on the number of ions~$N$, secular frequencies~$\omega_i'$ and normalized temperature~$T/\tau$.
We have compiled a two-dimensional map showing the scaling of the normalized critical temperature~$T_c/\tau$ as a function of~$N$ and trap anisotropy~$\alpha$ to quickly estimate hopping rates across a wide range of trapping parameters, found in Appendix~\ref{sec:heatmap}.

\section{Mixed-species linear ion crystals}
\label{sec:Mixed-species}

\begin{figure}
\centering
\includegraphics[width=\linewidth]{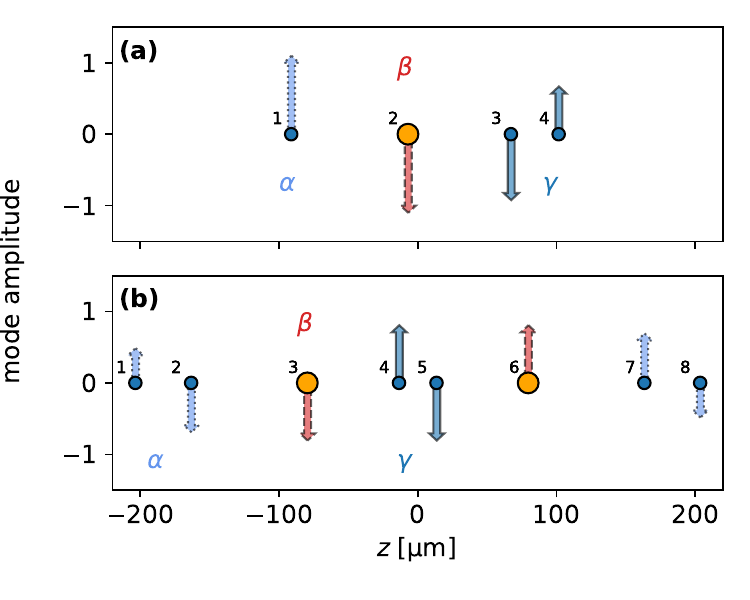}
\caption{Equilibrium ion positions along the $z$-axis for studied mixed-species configurations A~(a) and B~(b).
Small circles depict Be$^+$ ions, larger circles represent Ar$^{11+}$.
Secular frequencies of Be$^+$ ions in~(a) are $\omega_{x,y,z}'/2\pi = \{ 2.91, 3.60, 1.0 \} \times \SI{83.3}{\kilo\hertz}$ and for~(b) $\omega_{x,y,z}'/2\pi = \{ 10.1, 10.5, 1.0 \} \times \SI{54.3}{\kilo\hertz}$.
Three mode vectors $\alpha, \beta, \gamma$ of selected out-of-phase modes in $x$-direction are shown, differentiated by arrow style (amplitudes smaller than $0.02$ not depicted for clarity).
They display the projection of localized zigzag modes that trigger hopping on the radial motion of ions.
}
\label{fig:configs}
\end{figure}

\begin{figure*}
  \includegraphics[width=\textwidth]{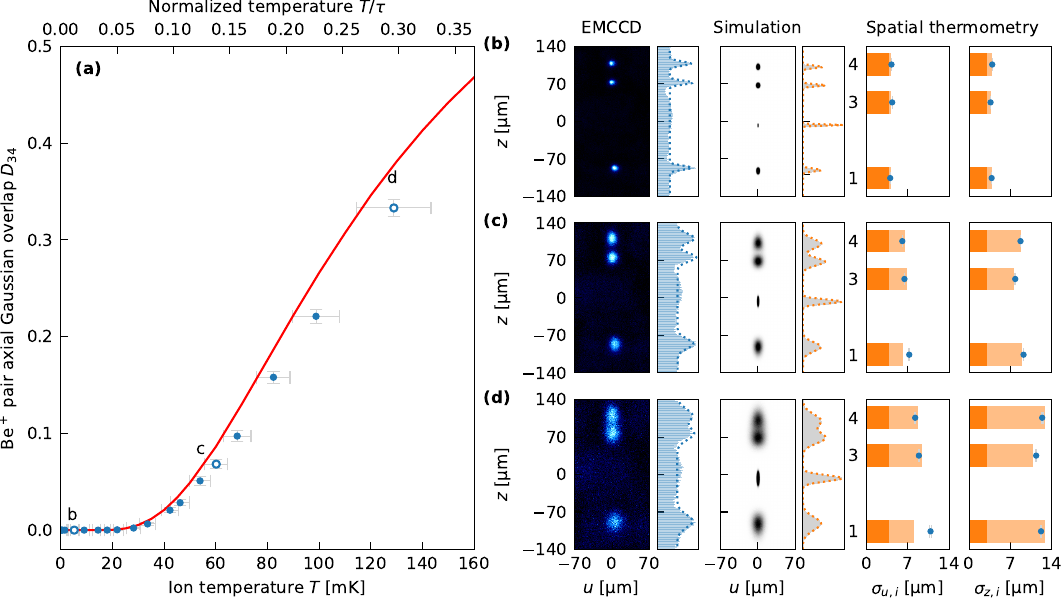}%
  \caption{%
  Average Gaussian overlap $D_{34}$ for a mixed-species crystal of three Be$^+$ ions and an Ar$^{11+}$ ion with $\omega_{x,y,z}'/2\pi = \{ 2.91, 3.60, 1.0 \} \times \SI{83.3}{\kilo\hertz}$. 
  The Gaussian overlap of just the two neighboring Be$^+$ ions (3, 4) (see Fig.~\ref{fig:configs}(a)) is shown instead of averaging over all pairs, as the overlaps involving the Ar$^{11+}$ ion remain $<10^{-4}$ over this temperature range. Experimental data are shown in (a) as blue points; simulation results as a solid red line.
  Insets (b)--(d) show data at three marked temperatures in greater detail: fitted \SI{25}{\second} exposure time EMCCD image, simulated data and spatial thermometry fit to determine~$T$ where the dark part of bars indicates fitted PSF contribution. Simulated scattering images include the position of the HCI which is not visible in the EMCCD images. The slight increase in scattering rate around the position of the HCI in the EMCCD images is caused by stray light from the cooling beam and unrelated to the HCI.
}
  \label{fig:3Be_1Ar_comparison}
\end{figure*}


We now extend our analysis to mixed-species ion crystals, where we continue to employ singly charged Be$^+$ ions as primary ion species and add Ar$^{11+}$ as secondary (HCI) species.
We focus on two specific crystal configurations, illustrated in Fig.~\ref{fig:configs}.
Configuration A is chosen for its ease of experimental realization, being the smallest crystal exhibiting the dynamics of interest for analyzing the influence of HCIs through both simulation and experiment.
Figure~\ref{fig:configs} also depicts the mode vectors of the localized zigzag modes, which we will discuss later.
In the case of configuration A, the trapping frequencies for singly charged Be$^+$ ions are chosen as $\omega_{x,y,z}'/2\pi \approx \{ 2.9, 3.6, 1.0 \} \times \SI{83.3}{\kilo\hertz}$.
The charge-to-mass ratio of $^{40}\text{Ar}^{11+}$ is around $2.5$ times higher than that of $^{9}\text{Be}^+$.
Using Equation~\eqref{eq:omegas} we find that the secular frequencies experienced by a single HCI for these trap parameters would be $\omega^\text{Ar}_{x,y,z} \approx \{ 2.7 \times \omega_x', 2.4 \times \omega_y', 1.6 \times \omega_z' \}$.

Configuration~B (Fig.~\ref{fig:configs}(b)) presents a more complex arrangement of Be$^+$ ions, segregated in three domains by argon ions.
For this configuration, the trapping frequencies for the singly charged species are $\omega_{x,y,z}'/2\pi \approx \{ 10.2, 10.5, 1.0 \} \times \SI{54.3}{\kilo\hertz}$ and for the HCIs $\omega^\text{Ar}_{x,y,z} \approx \{2.5 \times \omega_x', 2.5 \times \omega_y', 1.6 \times \omega_z' \}$. 

We will now present our experimental results of the reordering dynamics of crystal configuration A, before using the simulations to analyze the dynamics for both configurations in more detail.

\subsection{Comparison with experimental data}

Figure~\ref{fig:3Be_1Ar_comparison} presents experimental and simulated results of the overlap integral $D_{ij}$ for the two-Be$^+$ ion group (ions 3 and 4 in Fig.~\ref{fig:configs}(a)) of configuration A, applying the same analysis as described in Sec.~\ref{subsec:Single-species_Experimental_comparison}.
In the experiment, the HCIs do not scatter light from the Doppler cooling scheme as their electronic transitions are not resonant with these laser frequencies, see Fig.~\ref{fig:3Be_1Ar_comparison}(b)--(d).
It is therefore not possible to obtain results for the position and motion of the HCIs directly.
However, we can infer their position from the distribution of the bright Be$^+$ ions. Hopping of the HCI is immediately evident with the employed detection scheme.
We note that the position of the HCI in the ion chain remains stable even at the highest temperature set here by the Doppler laser detuning, whereas the group of two Be$^+$ (ions 3, 4) shows axial overlap and hopping already at $T/\tau \approx 0.05$.
This illustrates how the presence of the HCI greatly alters the relative hopping probabilities of the ion pairs compared to the four-ion single species crystal from Sec.~\ref{sec:Single-species}.
Fits of data and simulation are shown in more detail for selected data points in Figs.~\ref{fig:3Be_1Ar_comparison}(b)--(d).

The experimental data agrees well with the trend of the simulation results.
We find a small offset between the two, similar to the single-species case in Sec.~\ref{subsec:Single-species_Experimental_comparison}, due to issues of the imaging system. One additional effect not taken into account by the simulation is the light force imparted on the Be$^+$ ions by scattered photons from the cooling lasers.
Due to the small mass of Be$^+$, this shifts the crystal center by up to several \si{\micro\meter}, scaling linearly with the fluorescence rate.
Since HCIs do not scatter light from the Be$^+$ cooling lasers, the force only applies to Be$^+$ ions in the mixed-species crystal and somewhat distorts the linear configuration, plausibly causing the deviation in the radial spot size for ion~1 observed at higher temperature.
The qualitative agreement found provides a robustness check for the simulation methodology.
We will now further analyze the dynamics of configuration A and B using simulations.


\subsection{Localized melting}
\label{subsec:mixed_species_simulations}

\begin{figure*}[!htb]
\centering
 \includegraphics[width=\linewidth]{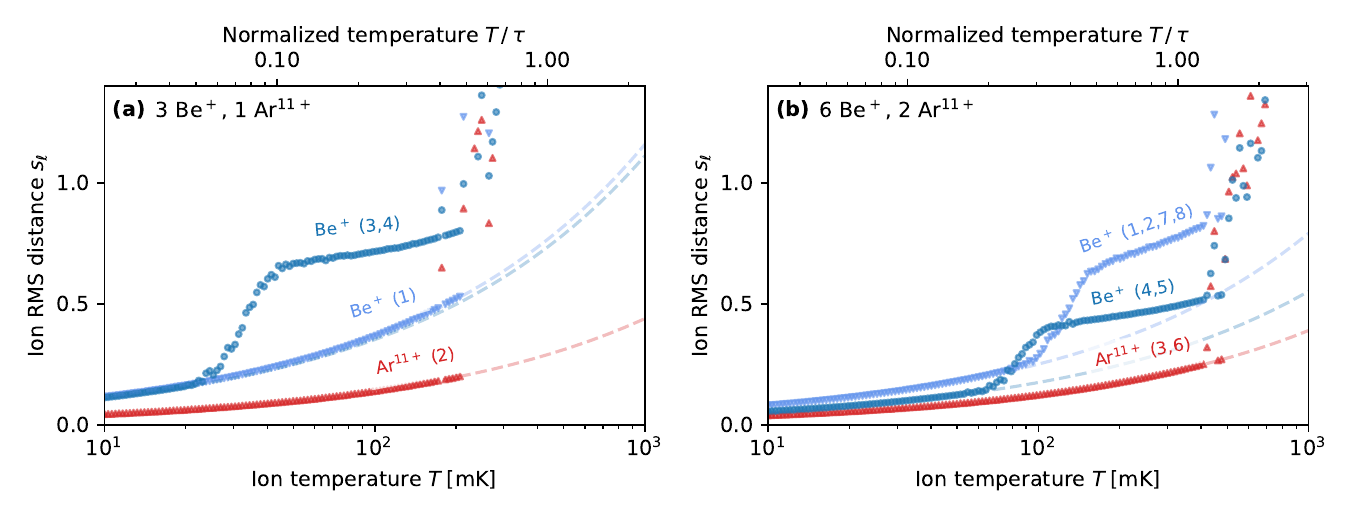}
\caption{Localized melting, demonstrated by the size of the ion oscillations for the crystal configurations A (Fig.~\ref{fig:configs}(a)) and B (Fig.~\ref{fig:configs}(b)). The root-mean-square displacement is plotted for the different crystal domains. Indices of the ions that were grouped together for the calculation of $s_\ell$ are shown above the respective data curves, labeled as in Fig.~\ref{fig:configs}. Data points depict results from numerical simulations while the dashed lines show the harmonic approximation. 
}
\label{fig:RMSD_mixed_species}
\end{figure*}

Similarly to our analysis of the single-species crystal, we plot the normalized root-mean-square displacement~$s_\ell$ by calculating Eq.~\eqref{eq:sigma} and plotting it as function of the temperature for both mixed-species crystals in Fig.~\ref{fig:RMSD_mixed_species}.
To illustrate the localized hopping dynamics within the different crystal domains formed by Be$^+$ ions separated by Ar$^{11+}$ ions, we calculate~$s_\ell$ for neighboring Be$^+$ ions and the Ar$^{11+}$ ions separately.
At low temperature, the mean amplitude of ion oscillations is reasonably well approximated by the harmonic theory depicted by dashed lines in Fig.~\ref{fig:RMSD_mixed_species}.
With increasing temperature, neighboring Be$^+$ ions hop between lattice sites, causing an increase in $s_\ell$ and deviations from the harmonic prediction. 
The temperature at which $s_\ell$ departs from the harmonic approximation depends on the product of the chosen simulated time and the hopping rate as already described in Sec.~\ref{subsec:Single-species_simulations}.
However, the relative temperature separation of the different hopping regimes is independent of the simulated time. 

In the low-temperature regime discussed so far, the Be$^+$ ions that exhibit hopping remain within their respective local group, such that their fluctuation amplitude is bound by the size of the respective Be$^+$ domain.
The HCIs act as fixed barrier between the different Be$^+$ domains.
Although we have not established a discrete melting transition from our results, we interpret the localized hopping dynamics as localized melting phenomena at different temperature regimes.
\begin{figure}
  \includegraphics[width=0.9\columnwidth]{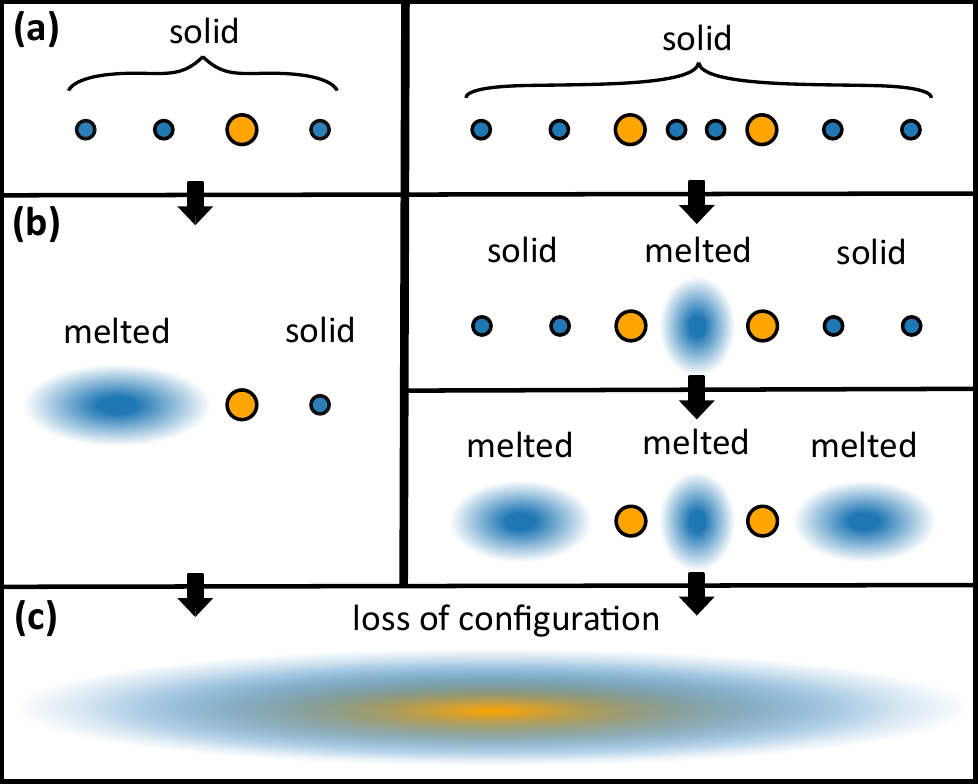}%
  \caption{Schematic representation of localized melting dynamics. Different regimes of increasing temperature are shown in (a)--(c). Left and right side show the four- and eight-ion mixed-species crystal of Fig.~\ref{fig:configs}. The different crystal domains, separated by HCIs melt locally at different temperatures, starting with the Be$^+$ domain that is closest to the trap center. (b) shows the coexistence of two phases at once (solid and liquid) within the same crystal. Once all domains of singly charged ions are melted and temperature increases further, the crystal loses its initial configuration due to the hopping of the HCIs (c).}
  \label{fig:localized_melting_schematic}
\end{figure}
In case of the four-ion mixed-species crystal (see Fig.~\ref{fig:RMSD_mixed_species}(a)), the thermal motion of the single Be$^+$ (ion~1) around its equilibrium position still matches harmonic theory (triangles pointing down) well into the temperature regime where another domain of the crystal has fully melted (circles).
Similar behavior can be seen in the eight-ion mixed-species crystal (Fig.~\ref{fig:RMSD_mixed_species}(b)), where $s_\ell$~of the inner two ions (circles) starts deviating from the harmonic approximation around $\SI{60}{\milli\kelvin}$, while for the outer ions it starts deviating at $\SI{90}{\milli\kelvin}$ (triangles pointing down).
Thus, at certain temperatures two structural phases can be present simultaneously within the same crystal in thermal equilibrium.
Figure~\ref{fig:localized_melting_schematic} illustrates this behavior schematically for the two mixed-species crystal configurations. 
We have already seen indications of these temperature regimes in the case of single-species crystals, where the inner ion pair starts hopping at lower temperature than ion pairs further away from the crystal center (Sec.~\ref{subsec:Single-species_simulations}).
While the symmetric eight-ion mixed-species crystal melts symmetrically, we find that the asymmetric four-ion mixed-species crystal also exhibits asymmetric melting behavior, with initial hopping not taking place between the inner ion pair.
Thus, adding Ar$^{11+}$ into the crystal does not only exaggerate the temperature separation of the local melting regimes, but also introduces an additional degree of freedom in shaping the inhomogeneity of the Coulomb crystal.

With increasing temperature, we observe the onset of Be--Ar hopping, indicated by a sharp jump in $s_\ell$.
At this point, the thermal energy of the particles has become comparable to the energy barrier for switching the $z$-ordering of a pair of Be$^+$ and Ar$^{11+}$ ions.
This energy scale is significantly higher than the Be$^+$--Be$^+$ barrier due to the larger Coulomb repulsion of the particles and the stronger confinement of HCI in the trap.
In the case of a single-species crystal, hopping of any ion pair switches between indistinguishable, degenerate configurations, but the hopping of an HCI can alter the crystal configuration and potential energy of the entire system.
We will now further analyze the dynamics of the HCI hopping based on the simulations.

\subsection{HCI hopping rates}
\label{subsec:mixed_species_hopping}

\begin{figure*}[!htb]
\centering
 \includegraphics[width=\linewidth]{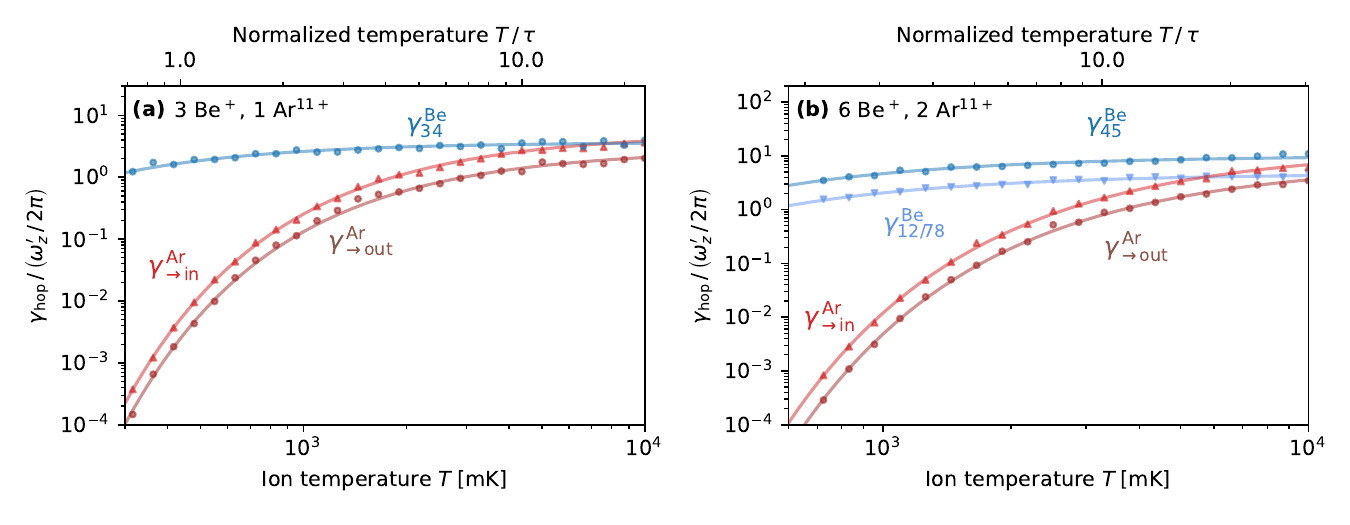}
\caption{Hopping rates of selected ion groups for the two crystal configurations (a) 3 Be$^+$, 1 Ar$^{11+}$ and (b) 6 Be$^+$, 2 Ar$^{11+}$ (see Fig.~\ref{fig:configs}).
For the Ar$^{11+}$, we plot the hopping rates which correspond to the HCI moving towards the crystal center ($\gamma^\text{Ar}_{\rightarrow\text{in}}$, triangles pointing up) and towards the crystal edges ($\gamma^\text{Ar}_{\rightarrow\text{out}}$, circles). The Arrhenius function is fitted to our data (solid lines). Secular frequencies of the Be$^+$ ions in (a) are $\omega_{x,y,z}'/2\pi = \{ 2.91, 3.60, 1.0 \} \times \SI{83.3}{\kilo\hertz}$ and in (b), $\omega_{x,y,z}'/2\pi = \{ 10.1, 10.5, 1.0 \} \times \SI{54.3}{\kilo\hertz}$. Error bars on the simulation data due to statistical fluctuations are not shown for clarity. However, it should be noted that all error bars are below $10\%$, ensuring that the data points are within an acceptable range of accuracy.}
\label{fig:hopping_mixed_species}
\end{figure*}

To further analyze the hopping dynamics of the HCIs, we plot the simulated hopping rates of the four-ion and eight-ion mixed-species crystal in Figs.~\ref{fig:hopping_mixed_species}(a) and (b), respectively.
The blue data depicts the hopping rates for neighboring Be$^+$ ions. Figure~\ref{fig:hopping_mixed_species}(b) has additional data in light blue, due to the extra pairs of Be$^+$ ions on the outside (see Fig.~\ref{fig:configs}(b)) which have identical hopping rates due to symmetry.
The indices of $\gamma^\text{Be}_i$ refer to the position in the crystal lattice shown in Fig.~\ref{fig:configs}.

In the case of the eight-ion mixed-species crystal, we find the hopping rate of the inner two Be$^+$ ions $\gamma^\text{Be}_{45}$ to be larger compared to the hopping rates of the outer Be$^+$ groups
$\gamma^\text{Be}_{12/78}$ as we would expect from the $s_\ell$ plots of Fig.~\ref{fig:RMSD_mixed_species}. 
To illustrate the anisotropy of the HCI hopping and thus the probability of finding the crystal in a certain configuration, we distinguish between HCI hops towards the crystal center $\gamma^\text{Ar}_{\rightarrow\text{in}}$ and hops towards the crystal edge $\gamma^\text{Ar}_{\rightarrow\text{out}}$.
For both configurations, the hopping rate towards the crystal center slightly outweighs the hopping rate towards the edge, since having the HCI close to the center minimizes the potential energy of the system. 
We use Eq.~\eqref{eq:exp_factor} to fit our data, taking $A$ and $\Delta E$ as free parameters.
We again find that the fit agrees reasonably well with our data within the depicted temperature range with slight deviations for $\gamma^\text{Ar}_{\rightarrow\text{out}}$ in the 4 ion crystal due to it having the least statistics.
Note that, due to the different trapping frequencies, the absolute values of $\gamma_\text{hop}/(\omega_z' / 2\pi)$ cannot be directly compared between the two crystal configurations.
While we here only consider the Ar$^{11+}$ hopping probability from the given starting configurations, additional simulations have shown that the hopping rate of Ar$^{11+}$ is dependent on its position inside the crystal, with hopping rates towards the crystal center always outweighing hopping rates towards the crystal edges.

From our fits, we find that the energy barrier extracted from the directional hopping rate is $\Delta E^\text{Ar}_{\rightarrow\text{in}} = \SI{260}{\micro\electronvolt}$ and $\Delta E^\text{Ar}_{\rightarrow\text{out}} = \SI{264}{\micro\electronvolt}$ for the four-ion mixed-species crystal (Fig.~\ref{fig:hopping_mixed_species}(a)).
The small difference in energy is the result of the trapping frequencies chosen and the small system size.
The energy barrier for the pair of Be$^+$ ions is a much lower $\Delta E^\text{Be}_{34} = \SI{29}{\micro\electronvolt}$.

In the case of the eight-ion mixed-species crystal (Fig.~\ref{fig:RMSD_mixed_species}(b)), we find {$\Delta E^\text{Ar}_{\rightarrow\text{in}} = \SI{608}{\micro\electronvolt}$} and {$\Delta E^\text{Ar}_{\rightarrow\text{out}} = \SI{632}{\micro\electronvolt}$} for the directional Ar$^{11+}$ hopping rates. For the inner Be$^+$ ions, we find {$\Delta E^\text{Be}_{45} = \SI{65}{\micro\electronvolt}$} and {$\Delta E^\text{Be}_{12/78} = \SI{71}{\micro\electronvolt}$} for the outer pairs.
Conclusively, hopping events where the HCI moves from the outside to the center of the crystal are statistically favored compared to HCI hops towards the edges.
For the eight-ion mixed-species crystal, any HCI hopping results in the loss of the initial configuration and as the two Ar$^{11+}$ ions preferably reside in the center of the chain, it rarely reappears.
The four-ion mixed-species crystal on the other hand is one of the two minimal energy states of that system.

\subsection{Locally decoupled modes}

The localized melting of the different Be$^+$ domains is also illustrated by the shape of the mode vectors in the most-weakly confined radial direction.
Figure~\ref{fig:configs} depicts three mode vectors $\alpha, \beta, \gamma$ of selected out-of-phase modes for each crystal configuration as differently styled arrows.
Comparing the mixed-species crystal of Fig.~\ref{fig:configs}(a) to its single-species counterpart of Fig.~\ref{fig:4_Be_zigzag_mode} shows that the projection of the local out-of-phase motion of ions 3 and 4 ($\gamma$, solid arrows) onto ion~1 is suppressed in the mixed-species case.
A similar observation can be made for the eight-ion mixed-species crystal, where the zigzag oscillation of the two inner Be$^+$ ions ($\gamma$, solid arrows) exhibits negligible influence on the zigzag motion of the outer Be$^+$ ion pairs, and vice versa. 
We find that the eigenfrequencies of the separated Be$^+$ domains at the edges of the eight-ion mixed-species crystal are degenerate for the localized zigzag mode as well as for localized common modes (ions within each group moving in the same direction).
This underlines the decoupling of the outer Be$^+$ pairs, essentially acting as isolated two-ion systems.
The much higher oscillation frequencies of the HCIs lead to pronounced decoupling of the argon ion motion from the more-slowly oscillating Be$^+$ ions.
In addition, its Coulomb repulsion increases the distance to neighboring groups of Be$^+$ ions.
As a result, HCIs shape the mode spectrum to exhibit localized modes that effectively partition the crystal into distinct vibrational domains.

The observed differences in eigenfrequencies for the local zigzag modes falls in line with the spatially different energy barriers (see Sec.~\ref{subsec:mixed_species_hopping}).
For example, in the case of the eight-ion mixed-species crystal, the frequency of the zigzag mode of the inner Be$^+$ ions is $\omega^{zz}_\text{in} = 9.01 \times \omega_z'$ and for the outer Be$^+$ ions $\omega^{zz}_\text{out} = 9.85 \times \omega_z'$, causing the energy needed for hopping to be reduced for the inner pair of Be$^+$.
In contrast, the eigenfrequency of the local zigzag mode of the HCIs ($\beta$, dashed arrows in Fig.~\ref{fig:configs}(b)) is $\omega^{zz}_\text{Ar} = 53.72 \times \omega_z'$, consistent with the increased energy barrier for the Be--Ar hopping.
Furthermore, when comparing a two-ion Be$^+$ crystal that has the same ion spacing and zigzag mode frequency as the two inner Be$^+$ ions in the eight-ion mixed-species crystal, we find that the critical temperatures are identical within the margin of error. 
Our findings indicate that the different Be$^+$ domains can be treated as isolated systems with dynamics that are decoupled from the rest of the crystal.

\section{Conclusion}
\label{sec:Conclusion}

In this paper, we have analyzed the thermal motion and reordering dynamics of both pure and mixed-species linear trapped ion crystals. By introducing Ar$^{11+}$ in a matrix of Be$^+$ crystals, we observe domain formation and structural super-lattices of different stabilities induced by the large charge-to-mass ratio of the highly charged ions. The temperature of pure and mixed species crystals can be reliably extracted experimentally from spatial thermometry which we benchmarked against simulations. 

The energy barrier between different permutations of the ions along their weakest confinement determines the onset of hopping. Therefore, the small size and inhomogeneity of the system make established melting criteria inapplicable. By modeling hopping rates as thermally-activated exponential processes, we can extrapolate reordering rates over a wide range of temperatures and trapping parameters. We investigated the scaling of thermal dynamics in single species systems, depending on the energy barrier and ion number and provide means for estimating the inner pair hopping rate for any single-species linear crystals of up to 20 ions.

By adding HCIs into the crystal, we obtain an extra degree of freedom in shaping the inhomogeneity of the crystal and demonstrate local melting effects.
Crystal `domains' formed by Be$^+$, disconnected by HCIs, melt at different temperatures depending on the size and configuration of the crystal.
These domains can be treated as separate Coulomb systems with local phonons that are decoupled from other parts of the crystal. 

Our work demonstrates the influence of HCIs on localized melting temperatures and the emergence of melting patterns imprinted in the shape and the frequencies of the crystal phonon modes.
At higher temperatures, reordering events involving HCIs become statistically significant, with the system tending towards states where the HCIs are located at the trap center due to the dependence of the axial trap potential on the ions charge-to-mass ratio.

Different charge-to-mass ratios alter coupling strengths between different crystal domains, providing even more degrees of freedom for the shaping of vibrational modes. Future research could lead to precise control over the thermal behavior of the crystal, advancing the study of phase transitions in mixed-species Coulomb crystals.

This paper is relevant to experiments with HCIs employing singly charged ions for sympathetic cooling by providing insights into the stability of crystal configurations against thermal fluctuations. The HCIs low susceptibility to external fields and their ability to induce distinct vibrational modes illustrate the potential for more controlled and tunable mixed-species Coulomb crystals in experiments involving quantum simulations or frequency metrology with high precision.

While our analysis is focused on one-dimensional crystals which are used in metrology applications, it motivates an extension to higher-dimensional structures.
In the case of two-dimensional ion lattices, the influence of the location of the HCIs onto the motional modes could emerge in intriguing patterns of melted domains co-existing with regions with negligible ion hopping.
Moreover, the effect of a string of HCIs surrounded by a three-dimensional cloud of singly charged ions onto the rotational dynamics of different ion shells and their mutual nanofriction could be investigated.

\begin{acknowledgments}

We would like to thank Ramil Nigmatullin for providing the basis of the simulation codes that were used during our research.

We acknowledge support by the projects 18SIB05 ROCIT and 20FUN01 TSCAC.
These projects have received funding from the EMPIR programme cofinanced by the Participating States and from the European Union’s Horizon 2020 research and innovation programme.
This project has been funded by the Deutsche Forschungsgemeinschaft (DFG, German Research Foundation) under Germany’s Excellence Strategy -- EXC-2123 QuantumFrontiers–390837967 and through CRC 1227 (DQ-mat), project A07.
This work has been supported by the Max Planck Society; the Max-Planck--Riken--PTB-Center for Time, Constants and Fundamental Symmetries; and the German Federal Ministry of Education and Research (BMBF) through program grant No. 13N15973 (Projekt VAUQSI).

\end{acknowledgments}

\bibliography{paper}

\appendix

\section{Molecular dynamics simulations}
\label{sec:simulations_methods}

At finite temperature~$T$, the ions do not remain in the static configuration~$\left\{\vec{r}_i^{\,0}\right\}$ but undergo stochastic trajectories.
We describe the dynamics using the Langevin formalism~\cite{Langevin1908}, which incorporates the Brownian motion of ions by introducing stochastic forces~$\vec\epsilon_i(t)$ in combination with a drag term parametrized by friction coefficient~$\eta$.
The resulting equations of motion take the form
\begin{align}
m_i\frac{d^2\vec r_i}{dt^2} = -\frac{d}{d\vec r_i}\mathcal{V} - m_i\eta\frac{d\vec r_i}{dt} + \vec \epsilon_i(t).\label{eq:Langevin}
\end{align}
To achieve thermalization of the ions at temperature~$T$, the stochastic forces are randomized, satisfying
\begin{equation}
    \expect{\vec{\epsilon}_i(t)} = \vec{0} \text{,} \quad \expect{\vec{\epsilon}_{i}(t) \otimes \vec{\epsilon}_{j}(t')} = 2 m_i \eta k_B T \delta_{i j} \delta(t-t') \mathbb{1}_3 \text{,}
\end{equation}
where $\otimes$~denotes the outer product and $\mathbb{1}_3$~is the $3 \times 3$ identity matrix.
In trapped-ion experiments, the friction and stochastic forces are caused by the absorption and spontaneous emission of photons from a cooling laser.
The friction coefficient~$\eta$ for a Doppler cooling scheme utilizing a transition at an optical wavelength~$\lambda$ has a maximal value at optimal detuning of $m_i \eta \approx \hbar \pi^2 / \lambda^2$~\cite{Leibfried2003}, which is typically on the order of \SIrange[print-unity-mantissa=false]{e-21}{e-20}{\kilogram\per\second}.

To investigate the non-equilibrium properties of the ion crystals depending on the temperature~$T$, we numerically solve the Langevin equations~\eqref{eq:Langevin} and subsequently build time averages of the quantities of interest~\cite{SkeelIzaguirre2002}.
For each choice of parameters, we first find the equilibrium positions~$\left\{\vec{r}_i^{\,0}\right\}$ by initializing the particles at random positions and simulating the resulting dynamics with $T=0$ until the kinetic energy of the ions reaches zero.
For simulations with~$T>0$, we then take the determined equilibrium configuration as the initial state and simulate the system for several times the timescale~$\eta^{-1}$ to reach thermalization at the given temperature.
The equipartition theorem can be employed to verify that the kinetic energy of the ions matches the desired thermal state.

After thermalization, we log the simulated ion positions and velocities to determine the quantities of interest.
The choice of time step and total simulated time needs to consider the discrepancy in oscillation periods between co-trapped singly charged ions and HCIs at identical trap parameters.
We find that choosing the integration time step about a factor~100 smaller than the fastest secular frequency of the simulated (mixed-species) crystal avoids numerical errors. Moreover, we make the run time sufficiently long to build meaningful time averages of ion motion.
The same simulation may then be repeated with different pseudo-random seeds to study the statistical distribution of results.


\section{Experimental setup}
\label{sec:appendix_experimental_setup}


To benchmark the simulation results, we compare the simulations with the dynamics observed of ions trapped in a linear Paul trap.
The cryogenic ion trap of the CryPTEx-SC experiment (see Fig.~\ref{fig:experiment_setup}) is of quasi-monolithic design detailed elsewhere~\cite{Stark2021, Dijck2023} operating at a frequency of \SI{35}{\mega\hertz} with blade RF electrodes spaced \SI{1.75}{\milli\meter} away from the trap center.
Eight DC electrodes with the same profile as the blade electrodes are set at an axial distance of \SI{2.05}{\milli\meter} from the center to provide axial confinement and compensate stray electric fields.
The trap is operated primarily with singly charged Be$^+$ ions, loaded by photo-ionizing $^9$Be atoms from a collimated beam produced by a resistively-heated oven mounted about \SI{1}{\meter} away from the trap center.
A series of apertures prevents contamination of trap surfaces by the atomic beam.

The experimental setup additionally provides for the loading of HCIs~\cite{Schmoeger2015, Dijck2023}.
These are extracted in pulses from an electron beam ion trap (EBIT) based on permanent magnets~\cite{Micke2018}, connected to the linear Paul trap via a transfer beamline.
In this work, the EBIT was operated with continuously injected argon gas.
The charge state of HCIs sent to the Paul trap is controlled by the breeding time between EBIT extractions of order \SI{100}{\milli\second} and the electron beam energy.
The transfer beamline comprises several ion optics for steering and focusing the HCIs, as well as a pulsed drift tube for bunching and initial deceleration of their kinetic energy from \SI{700}{\volt \times \mathit{Q}} after extraction to about \SI{150}{\volt \times \mathit{Q}}, where $Q$ is the charge state.
The kinetic energy of the HCIs is further reduced to only a few volt by operating the Paul trap at an elevated ground potential just below the HCI energy.
Switchable mirror electrodes at each end of the Paul trap RF blades are used to capture HCIs before final deceleration and co-crystallization in a pre-loaded Be$^+$ crystal.
The Paul trap is operated at a temperature of about \SI{5}{\kelvin} to bring down the background gas pressure and reduce the rate of HCI charge exchange reactions.

Doppler cooling is performed on the Be$^+$ $1\text{s}^2\;^2\text{S}_{1/2}$ -- $1\text{s}^2\;^2\text{P}_{3/2}$ transition by laser light at wavelength \SI{313}{\nano\meter}.
Operating with circular polarization and aligning the laser propagation direction along the quantization magnetic field of about \SI{200}{\micro\tesla} drives the $\text{S}_{1/2}$ ($F=2$, $m_F=+2$) -- $^2\text{P}_{3/2}$ ($F=3$, $m_F=+3$) cycling transition.
In general, imperfections in the polarization and laser pointing require some repumping from the upper $F=1$ ground state hyperfine level, which is implemented by a second laser at \SI{313}{\nano\meter} driving excitation to the $1\text{s}^2\;^2\text{P}_{1/2}$ level.
The wavelengths of the lasers are controlled by a wavemeter referenced to a stabilized helium-neon laser.
The laser beams and quantization axis are set at an angle of \SI{30}{\degree} to the trap center line, projecting onto all three principal trap axes to enable cooling of all motional modes of ion crystals.
When trapping HCIs, the Be$^+$ ions provide sympathetic cooling.

Ion fluorescence is captured by cryogenic imaging optics refocusing the light onto an EMCCD camera (electron-multiplying charge-coupled device) and a photomultiplier tube outside the vacuum chamber.
The optics consist of a lens stack mounted on top of the ion trap~\cite{Dijck2023} with a numerical aperture of about $\text{NA} \approx \num{0.36}$ producing an image with a magnification of about~$9.7 \times$.

Under optimal settings, the ions are cooled down close to the Doppler limit~\cite{Leibfried2003}
\begin{equation}
    T_D \approx \frac{\hbar \Gamma_D}{2 k_B} \text{,}
\end{equation}
where $\Gamma_D$ is the linewidth of the transition used for Doppler cooling, yielding \SI{0.5}{\milli\kelvin} in the case of Be$^+$.
This limit is achieved when the heating caused by spontaneous emission in the Doppler process is balanced by the cooling effect for vanishing laser intensity at a detuning of $2\pi \Delta\nu = - \Gamma_D / 2$.
Changing the intensity or detuning of the cooling laser light allows to shift the point where this balance is reached to higher temperatures.
We use this control of ion temperature to study thermal dynamics in the experiment.

\section{Harmonic approximation}
\label{sec:analytic_msd}

Here we sketch the calculation of the results for the spatial extent of the ion oscillations in harmonic approximation.
For small energies such that the displacements of the ions from their equilibrium position at $T=0$, denoted by $\vec{q}_i=\vec{r}_i-\vec{r}_i^{\,0}$, are small in comparison to the inter ion spacing $\left|\vec{r}_i^{\,0}-\vec{r}_j^{\,0}\right|$ we can expand the potential energy Eq.~\eqref{eq:potential} up to second order.
This yields the dynamical matrix $\mathcal{K}$ with blocks
\begin{equation}
    \mathcal{K}_{ij}=\frac{\partial^2 \mathcal{V}}{\partial \vec{r}_i \partial \vec{r}_j}\big|_{\vec{ r}_i^{\,0}}
\end{equation}
and eigenvectors $\vec v^{\,\mu}$ and eigenvalues $\lambda_\mu$.
The displacements of the ions can be written in the form
\begin{equation}
    \vec q_i = \sum_\mu \vec v_i^{\,\mu} \Theta_\mu
\end{equation}
with the mode amplitude lengths~$\Theta_\mu$.

The extent of the thermal fluctuations of the ions can be calculated as
\begin{align}
    \label{eq:mean_displacement}
    \left\langle \vec q_i^{~2} \right\rangle &= \sum_{\mu\nu} \vec v_i^{\,\mu} \cdot \vec v_i^{\,\nu} \left\langle \Theta_\mu \Theta_\nu \right\rangle \\
    &= \sum_\mu \left(\vec v_i^{\,\mu}\right)^2\left\langle\Theta_\mu^2\right\rangle
\end{align}
where we have assumed for the last line that the spectrum of $\mathcal{K}$ is free of degeneracies.
Under the presumption of equal coupling of all ions to the Langevin heat bath we can relate the average energy of each mode to the reservoir temperature by the equipartition theorem
\begin{equation}
    \label{eq:equipartition}
    \frac{1}{2} \lambda_\mu \left\langle \Theta^2_\mu \right\rangle = \frac{1}{2} k_B T.
\end{equation}
Inserting into the squared displacement gives 
\begin{equation} 
\label{eq:harmonic_msd}
    \left\langle \vec q_i^{~2} \right\rangle = k_B T \sum_{\mu} \frac{\left( \vec{v}_i^{\,\mu} \right)^2}{\lambda_\mu} \text{.}
\end{equation}

\section{Discussion on the Lindemann parameter and the \texorpdfstring{$\Gamma$}{Gamma} melting criterion}
\label{sec:Lindemann_Gamma_discussion}

\begin{figure}
  \includegraphics[width=\columnwidth]{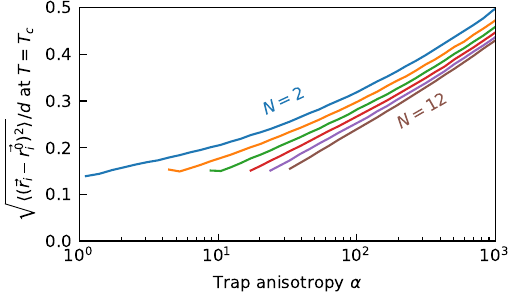}
  \caption{%
Average RMS normalized by the average ion spacing at the critical temperature $T=T_c$ for single-species strings as function of trap anisotropy~$\alpha$ for ion number $N=2,4,\ldots,12$.
}
  \label{fig:lindemann_vs_alpha}
\end{figure}

\begin{figure}
\includegraphics[width=\columnwidth]{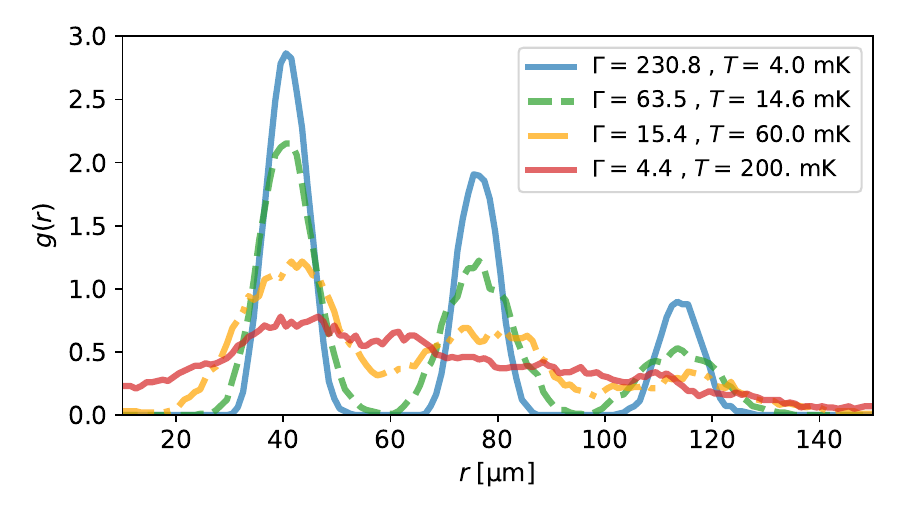}
\caption{One-dimensional spatial correlation function $g(r)$ for the four-ion Be$^+$ crystal at different temperatures. The corresponding $\Gamma$ parameter is indicated in the key.
}
\label{fig:4_Be_correlation}
\end{figure}

\begin{figure*}
  \centering
  \includegraphics[width=0.7\textwidth]{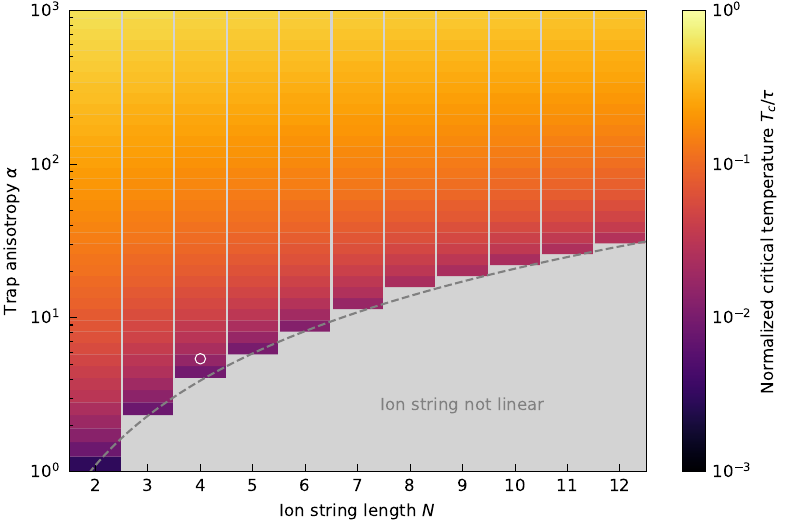}
  \caption{Scaling of the normalized critical temperature~$T_c/\tau$ for simulated single-species ion strings of up to $N=12$ ions as function of axial--radial trap anisotropy $\alpha = \omega_x'^2 / \omega_z'^2$. Linear strings require a minimum value of~$\alpha$ that depends on~$N$, below which the ions form a zigzag configuration instead; the dashed line shows an approximate expression for this critical anisotropy~\cite{Enzer2000}. The white circle shows the parameters used in the experiment and simulations of Sec.~\ref{sec:Single-species}.}
  \label{fig:Heatmap}
\end{figure*}

While our critical threshold for the hopping rate is motivated by experimental considerations, this approach based on the rate of crystal reorderings illustrates the dynamics more adequately than the Lindemann melting criterion.
To demonstrate this, we plot in Fig.~\ref{fig:lindemann_vs_alpha} the root-mean-square displacement (RMS) divided by the average equilibrium ion spacing (Eq.~\eqref{eq:Lindemann}) at the critical temperature~$T_c$ determined from the hopping rate as a function of trap anisotropy~$\alpha$.
We find that the onset of hopping coincides with a Lindemann threshold~$C_L$ of $\approx 0.14$ in the low trap anisotropy regime, however, this result evidently depends on the chosen threshold value for the hopping rate.
The graph furthermore shows that for increasing trap anisotropy~$\alpha$, the corresponding $C_L$ grows such that no unique Lindemann threshold can be found for the onset of hopping which we consider the manifestation of the crystal melting.
An $\alpha$-dependent threshold $C_L$ could be considered, however, its functional dependence is a priori unclear and requires to consult other dynamical measures such as the hopping rate.
We note that similar arguments hold for the overlap $D_{ij}$ (see Sec.~\ref{subsec:Single-species_Experimental_comparison}).


Revisiting the $\Gamma$ melting criterion and assessing its validity for the four-ion crystal, we calculate the one-dimensional pair-correlation function $g(r)$, as well as the corresponding $\Gamma$-parameter at different temperatures. The pair-correlation function $g(r)$ gives a measure of the probability of finding a particle at distance $r$ of a reference particle. Oscillations in $g(r)$ show the spatial correlations between particles and disappear at high temperature. We plot $g(r)$ for the four-Be$^+$ ion crystal in Fig.~\ref{fig:4_Be_correlation} at different temperatures. The given $\Gamma$-parameter was calculated using Eq.~\eqref{eq:Gamma}, while the Wigner-Seitz radius was adjusted to the one-dimensional case being half of the mean inter-atomic distance at the simulated temperature. While the pair correlation function provides a measure of the internal structure of the system, it does not indicate a well-defined melting point for the small four-particle system, due to thermal fluctuations and pronounced effects of the finite size of the system. These factors lead to a continuous transition from ordered to disordered state, as the individual particle movement gradually increases with temperature, rather than showing a sudden shift characteristic of bulk systems. Using the critical temperature as defined in Section~\ref{subsec:Melting_Criteria}, we find that $\Gamma_m = 63.5$ (green dashed line). At that point, the oscillations in $g(r)$ remain noticeable, albeit with less pronounced peaks.

Neither the Lindemann nor the $\Gamma$-melting criterion reveal a discrete melting transition for small system sizes, as both criteria are based on variables that change continuously with temperature here. Although establishing a critical temperature $T_c$ based on a critical hopping rate is inherently arbitrary, it provides a threshold that does not require large isotropic systems to be applied and is especially relevant from an experimental perspective, as the reordering of linear mixed-species crystals can have significant implications on  experimental outcomes.

\section{Critical temperature look-up map}
\label{sec:heatmap}

We compiled an overview of the scaling of ion thermal dynamics in Figure~\ref{fig:Heatmap} by plotting the normalized critical temperature~$T_c/\tau$ for single-species linear ion crystals as a function of trap anisotropy~$\alpha$ (taking $\omega_x' = \omega_y'$) and the number of ions up to~$N=12$.
Note that forming a linear ion string requires a minimum trap anisotropy~$\alpha$ that depends on $N$.
The two-dimensional map allows to read off the temperature above which the hopping rate exceeds $\gamma_\text{hop} = 10^{-7} \times \omega_z'/2\pi$ as defined in Sec.~\ref{subsec:Scaling}, using Eq.~\eqref{eq:rescaled_length} to convert the normalized temperature into an absolute value.
The hopping rate (of the innermost ion pair) at different ion temperatures for these trap parameters can then be estimated by combining Eqs.~\eqref{eq:energy_barrier_estimation} and \eqref{eq:exp_factor}.

\end{document}